\documentclass[12pt,preprint]{aastex}
\notetoeditor{refer to: Fiamma Capitanio. Institute: IASF-INAF,Via Fosso del Cavaliere, 100, 00133 Rome ITALY. Tel: +390649934557. E-mail: fiamma.capitanio@rm.iasf.cnr.it}
\newcommand{\inst}{\altaffilmark}

\usepackage{graphicx}
\usepackage{natbib}
\shorttitle{IGR J17091--3624 and IGR J17098--3628: a multiwavelength long term campaign}
\shortauthors{Capitanio F. et al.}
\begin{document}
   \title{The two INTEGRAL X-ray transients IGR J17091--3624 and IGR J17098--3628: a multi-wavelength long term campaign}

\author {F. Capitanio\inst{1}, M. Giroletti\inst{2}, M. Molina\inst{3},
A. Bazzano\inst{1}, A. Tarana\inst{1}, J. Kennea\inst{4}, A. J. Dean\inst{3},
A. B. Hill\inst{3}, M. Tavani\inst{1}, P. Ubertini\inst{1}}
 
\altaffiltext{1}{INAF IASF-Roma, Via Fosso del Cavaliere 100, 00033 Rome,
Italy}

\altaffiltext{2}{INAF Istituto di Radioastronomia, via Gobetti 101, 40129
Bologna, Italy}

\altaffiltext{3}{School of Physics and Astronomy, University of Southampton,
Highfield Southampton, SO17 1BJ, UK}

\altaffiltext{4}{Department of Astronomy and Astrophysics, Pennsylvania State
University, University Park, PA 16802, Pennsylvania, USA}


\begin{abstract}

IGR J17091--3624 and IGR J17098--3628 are two X-ray transients discovered by
{\it INTEGRAL} and classified as possible black hole candidates (BHCs). We
present here the results obtained from the analysis of multi-wavelength data
sets collected by different instruments from 2005 until the end of 2007 on both
sources. IGR J17098--3628 has been regularly detected by {\it INTEGRAL} and
{\it RXTE} over the entire period of the observational campaign; it was also
observed with pointed observations by {\it XMM} and {\it Swift}/XRT in 2005 and
2006 and exhibited flux variations not linked with the change of any
particular spectral features. IGR J17091--3624 was initially in quiescence
(after a period of activity between 2003 April and 2004 April) and it was then
detected again in outburst in the XRT field of view during a {\it Swift}
observation of IGR J17098--3628 on 2007 July 9. The observations during
quiescence provide an upper limit to the 0.2-10 keV luminosity, while the
observations in outburst cover the transition from the hard to the soft
state. Moreover, we obtain a refined X-ray position for IGR J17091--3624 from
the {\it Swift}/XRT observations during the outburst in 2007. The new position
is inconsistent with the previously proposed radio counterpart. We identify in
VLA archive data a compact radio source consistent with the new X-ray position
and propose it as the radio counterpart of the X-ray transient.

\end{abstract}


\section{Introduction}

Among the sources included in the {\it INTEGRAL} IBIS/ISGRI survey
catalog~\citep{Bird}, the two X-ray transients IGR J17091--3624 and IGR
J17098--3628 are remarkable because of their proximity, being only 9\arcmin.6
away from each other.  The two sources have been detected in different periods
of time from April 2003 until the beginning of 2008. In particular, IGR
J17091--3624 was detected by IBIS in 2003 and remained detectable for one year
and, after a period of quiescence, it was again in outburst in July 2007. IGR
J17098--3628 was detected for the first time by {\it INTEGRAL} in 2005 (while
IGR J17091--3624 was not visible) and it has then remained detectable with
variable flux in the soft X-ray energy range (E $<$ 20 keV) up to now. On the
basis of their spectral behavior, both IGR J17091--3624 and IGR J17098--3628
are classified as probable BHCs \citep{Luto03,Greb2007}.

In this paper, we present the results of a long term monitoring, primarily at
high energy, of these two transients, and discuss the identification in
archival data of a radio counterpart for IGR J17091--3624. 
The date, the exposure time, the instruments used and the sources detection of the different observations are summarized in Table~\ref{log}.
The paper is organized as follows: in \S\ref{intro1} and \S\ref{intro2}, we introduce the
two BHCs; in \S\ref{obs}, we describe our observations; in \S\ref{res}, we
present the results; finally, we give a discussion and present our conclusions
in \S\ref{disc}.

\begin{table}
\begin{center}
\caption{IGR J17098-3628 and IGR J17091-3624 observations log from 2003 until September 2007}
\label{log}
\begin{tabular}{lcccc}
\hline
\hline

Start date & End date & Satellite & Detected sources & Exposure \\ 
2003-04-12 & 2003-04-21 & {\it INTEGRAL} & IGR J17091-3624\tablenotemark{a} & 15 ks \\
2003-04-20 & \nodata & {\it RXTE} & IGR J17091-3624\tablenotemark{b} & 4.5 ks  \\
2003-04-23 & 2003-05-09 & VLA\tablenotemark{c}& IGR J17091/IGR J17098 & 500 s \\
2003-08-10 & 2004-04-02 & {\it INTEGRAL} & IGR J17091-3624\tablenotemark{a} & 154 ks  \\
2005-02-20 & 2005-05-01 & {\it INTEGRAL} & IGR J17098-3628\tablenotemark{d} & 300 ks \\
2005-03-29 & \nodata & {\it RXTE} & IGR J17098-3628\tablenotemark{d}& 2.1 ks \\
2005-05-01 & 2007-02-28 & {\it INTEGRAL} & IGR J17098-3628 &  450 ks\\
2005-05-01 & 2005-09-14 & {\it Swift} & IGR J17098-3628 & 8.5 ks \\
2006-08-25 & \nodata & {\it XMM} & IGR J17098-3628  & 8 ks \\
2007-02-19 & \nodata  & {\it XMM} & IGR J17098-3628  & 16 ks \\
2007-07-19 & 2007-08-07 & {\it Swift} & IGR J17091/IGR J17098 & 11 ks \\
2007-08-25 & 2007-09-30 & {\it INTEGRAL} & IGR J17091-3624 & 50 ks  \\
\hline
\end{tabular}
\tablenotetext{a}{Capitanio et al. 2006}
\tablenotetext{b}{$ $Lutovinov \& Revnivtsev 2003}
\tablenotetext{c}{Rupen et al. 2003}
\tablenotetext{d}{Grebenev et al. 2007}
\end{center}
\end{table}

\subsection{IGR J17091--3624} \label{intro1}

IGR J17091--3624 was discovered by {\it INTEGRAL}/IBIS during a Galactic
Center observation on 2003 April 14--15~\citep{Kuulk}.  Initially, the flux was
$\sim$20 mCrab in the 40--100 keV energy band exhibiting a hard spectrum, while
it was not detected in the 15--40 keV band, with an upper limit of $\sim$10
 mCrab. During subsequent observations of the Galactic Center Deep Exposure
(GCDE) on 2003 April 15--16, the source flux increased to $\sim$40 mCrab in the
40--100 keV band and to 25 mCrab in the 15--40 keV (the IBIS flux statistical
error is less then 10\%).

Immediately after the {\it INTEGRAL} discovery, providing the position of 
IGR J17091--3624, an {\it RXTE\/} observation was performed and the source
 was then searched in the X-ray catalogs. IGR J17091-3624 was found in the 
archival data of both the TTM telescope on board the KVANT module of the 
{\it Mir\/} orbital station \citep{Atel2}, and in the {\it BeppoSAX} WFC \citep{Atel4}.
A first study of the IBIS/ISGRI spectral evolution of the source~\citep{Luto03, Luto05}
showed a source hardening with a photon index changing from $\Gamma = 2.2 \pm 0.1$ to
$\Gamma=1.6\pm 0.1$ from 2003 April to 2003 August. A subsequent detailed
analysis of the IBIS, JEM-X and {\it RXTE}/PCA data of the entire outburst
duration~\citep{Cap2006} revealed an indication of an hysteresis like behavior
and the presence of a hot disc black body emission component during the source
softening.

From the investigations reported above, IGR J17091--3624 appears as a
moderately bright variable transient source, with a flaring activity in 1994
October ({\it Mir}/KVANT/TTM), 1996 September ({\it BeppoSAX}/WFC), 2001
September \citep[{\it BeppoSAX}/WFC,][]{Atel4}, 2003 April \citep[{\it
INTEGRAL}/IBIS,][]{Kuulk}, and 2007 July. In this paper, we report on the last
episode of activity, as well as on limits on the quiescent state.

\subsection{IGR J17098--3628}  \label{intro2}
 
IGR J17098--3628 was detected for the first time with {\it INTEGRAL}/IBIS 9.4'
off IGR J17091--3624 \citep{Atel444} during deep Open Program observations of
the Galactic Center region on 2005 March 24.  The average fluxes were 28.2
$\pm$ 1.4 and 38.7 $\pm$ 2.8 mCrab in the 18--45 and 45--80 keV bands,
respectively.  Further analysis \citep{Atel447} reported that the source was
evolving in both brightness and spectral shape with an indication of
softening.

From 2005 March 29 to April 4, an observational campaign was performed with {\it
RXTE}/PCA. The spectral shape given by {\it INTEGRAL}/IBIS and {\it RXTE}/PCA
varied throughout the observations and it is modelled by a soft black body
emission component plus a hard tail and an absorption consistent with the
Galactic one~\citep{Greb2007}. The spectral variations suggested that this
source was an X-ray nova going in outburst and probably a BHC \citep{Greb2007}.

The source was then observed with the {\it Swift} satellite \citep{Atel476}
with an exposure time of 2.8 ks on 2005 May 1; it was quite bright, with
an estimated flux of 1.3 $\times$ $ 10^{-9}$ ergs s$^{-1}$ cm$^{-2}$ in the 0.5--10 keV
energy band.  The analysis of the {\it Swift}/XRT data for IGR J17098--3628
refined the source coordinates as follows: RA 17$^h$ 09$^m$ 45.9$^s$, Dec --36$^\circ$
27$\arcmin$ 57$\arcsec$ (J2000), with an uncertainty radius of about 5$\arcsec$
\citep{Atel476}. This position is 30$\arcsec$ from the {\it INTEGRAL} position
reported by \cite{Atel444}.


Following the soft X-ray detection of IGR J17098--3628 with {\it Swift}/XRT, a
probable radio counterpart has also been found \citep{Atel490}. In particular,
the first data set of four consecutive Very Large Array (VLA) radio
observations, made on 2005 March 31, April 5, April 12, and May 4 at 4.86 GHz,
showed only one significant radio source within the 2$\arcmin$ {\it INTEGRAL}
error circle, located at RA 17$^h$ 09$^m$ 45.934$^s$ $\pm$ 0.011s, Dec --36$^\circ$
27$\arcmin$ 57.30$\arcsec~\pm$ 0.55$\arcsec$ (J2000) \citep{Atel490}.

Thanks to the radio observations, a probable optical/infrared identification
has been found within the 2MASS All-Sky Catalog and the SuperCOSMOS Sky
Survey~\citep{Atel478,Atel479,Atel490,Atel494}.  On the base of the optical
identification, \citet{Greb2007} estimated the source distance as ${\it d}$=
10.5 kpc and an upper limit on the inclination angle: ${\it }\geq$
77$^{\circ}$.

\section{The multi-wavelength observation campaigns}  \label{obs}

\begin{figure}
  \plotone{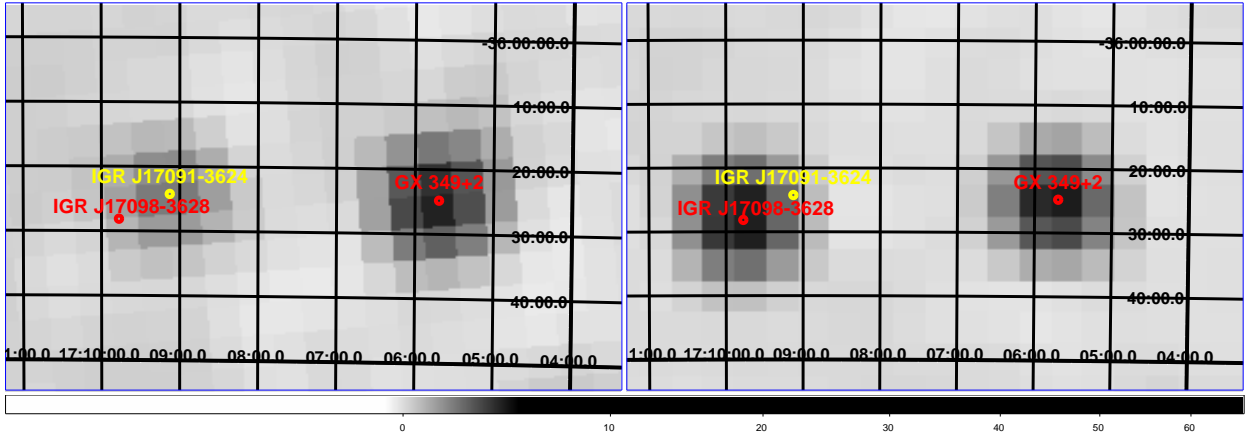}
  \caption{Left: 2003 IBIS/{\it INTEGRAL} 80 ks mosaic image (20-40 keV) of IGR J17091--3624 and IGR J17098--3628 zone: only IGR J17091--3624 is visible. Right: 2005 IBIS/{\it INTEGRAL} 120 ks mosaic image (20-40 keV) of IGR J17091--3624 and IGR J17098--3628 zone: only IGR J17098--3628 is visible}
  \label{2003-2005IBIS}
 \end{figure}

We report here on the analyzed data collected by {\it INTEGRAL}, {\it Swift}
and {\it XMM} from 2005 till 2007. To illustrate the need for a multi-wavelength
campaign with good positional accuracy, we show in Figure~\ref{2003-2005IBIS} two
{\it INTEGRAL} mosaic images of the region of the two BHCs. In the left panel,
we show the region during the IGR J17091--3624 outburst in 2003 (IGR
J17098--3628 was switched off), while in the right panel we show the same zone
when only IGR J17098--3628 was visible.

\begin{figure}
\plotone{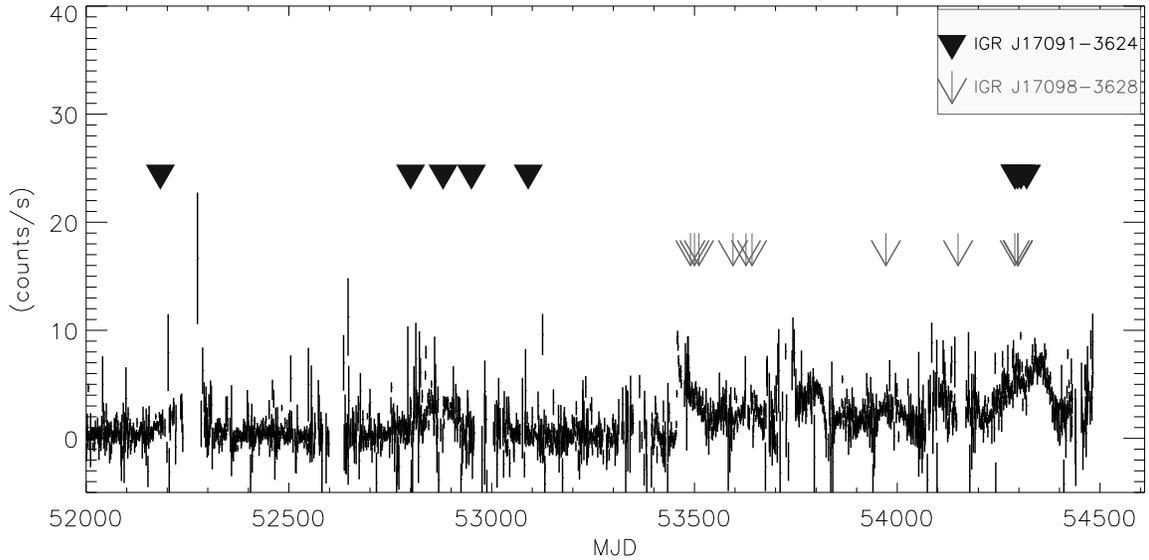}
      \caption{IGR J17098--3628 {\it RXTE}/ASM light curve contaminated by IGR J17091--3624 outbursts from 2001 till January 2008. The black triangles and gray arrows represent the periods of {\it BeppoSAX}, {\it Swift}/XRT and {\it XMM} observation campaigns that verified the activity of respectively IGR J17091--3624 and IGR J17098--3628. The time is expressed in Modified Julian Date (52000 (MJD)= 2001-04-01; 53500 (MJD)= 2005-05-10; 54500 (MJD)= 2008-02-04).}
         \label{2007ASM}
 \end{figure}

The two sources are too close to each other to be resolved by X-ray and
$\gamma$-ray monitors such us {\it RXTE}/ASM and {\it Swift}/BAT. Therefore, it
is not possible to measure their flux separately and it is rather difficult to
follow their individual activities. Observations with better positional
accuracy are needed to assign each flux variation to one of the two sources
with confidence. As an example, Figure~\ref{2007ASM} shows the contaminated
{\it RXTE}/ASM IGR J17098--3628 plus IGR J17091--3624 light curve between
1.5--12 keV from 2001 until the beginning of 2008. The black and gray arrows
represent the periods of {\it BeppoSAX}, {\it Swift}/XRT and {\it XMM} observation campaigns that
verified the activity of respectively IGR J17091--3624 and IGR J17098--3628.

The X-ray Telescope (XRT) on board {\it Swift} satellite is an imaging
instrument operating in the 0.2--10~keV energy range with a single photon point
spread function of 18$\arcsec$ (half-power diameter) and a spectral resolution
of 140~eV at 5.9~keV. We have collected all the XRT data available from the
NASA HEASARC public
archive\footnote{http://heasarc.gsfc.nasa.gov/docs/archive.html}. The XRT data
were processed using the most recently available standard {\it Swift} tools:
XRT software version 0.11.4, FTOOLS version 6.3.1, and XSPEC version 11.3.2.
The ancillary response files were generated with the {\it xrtpipeline} task
{\it xrtmkarf}. The channels of each spectrum were re-binned in order to achieve
a minimum of 100 counts per bin.  The {\it Swift}/BAT transient monitor light
curve was provided by the {\it Swift}/BAT
team\footnote{http://swift.gsfc.nasa.gov/docs/swift/results/transients/index.html}.

We also analyzed two {\it XMM} pointed observations performed on 2006 August
25 and 2007 February 19. {\it XMM} data were reprocessed using the {\it
XMM}-Newton Standard Analysis Software (SAS) version 7.0, employing the latest
available calibration files.  Only single X-ray events (PATTERN=0) were taken
into account for the PN;
the standard selection filter FLAG=0 was applied. Exposures have been filtered
for periods of high background. Since pile-up was present in both observations,
source counts were extracted from annular regions of typically 90$\arcsec$
external radius centered on the source, and the central 7$\arcsec$ of the PSF
have been excised; background spectra were extracted from circular regions
close to the source or from source-free regions of typically 20$\arcsec$ of
radius. The ancillary response matrices (ARFs) and the detector response
matrices (RMs) were generated using the {\it XMM}-SAS tasks {\it arfgen} and
{\it rmfgen}; spectral channels were re-binned in order to achieve a minimum of
100 counts per each bin.

The analyzed {\it INTEGRAL} data set consists of all Key Program and public
observations from 2005 May until the end of 2007. The {\it INTEGRAL} data
reduction of both the X-ray monitor JEM-X~\citep{Lund} and the $\gamma$-ray
telescope IBIS~\citep{Ube} were performed using the latest release of the
standard Offline Scientific Analysis (OSA)~\citep{Gold} version 7.

We have also searched in the radio archives for data containing IGR J17091--3624
and IGR J17098--3628 observations. Very Large Array (VLA)\footnote{The National
Radio Astronomy Observatory is operated by Associated Universities, Inc., under
cooperative agreement with the National Science Foundation.} observations were
taken for a few epochs in 2003, following the outburst of IGR J17091--3624, and
in 2005, after the IGR J17098--3628 detection. We have obtained and calibrated
the data in AIPS with the standard procedures.

\section{Results} \label{res}
\subsection{Swift 2005 Observations}
 
\begin{figure}
\plotone{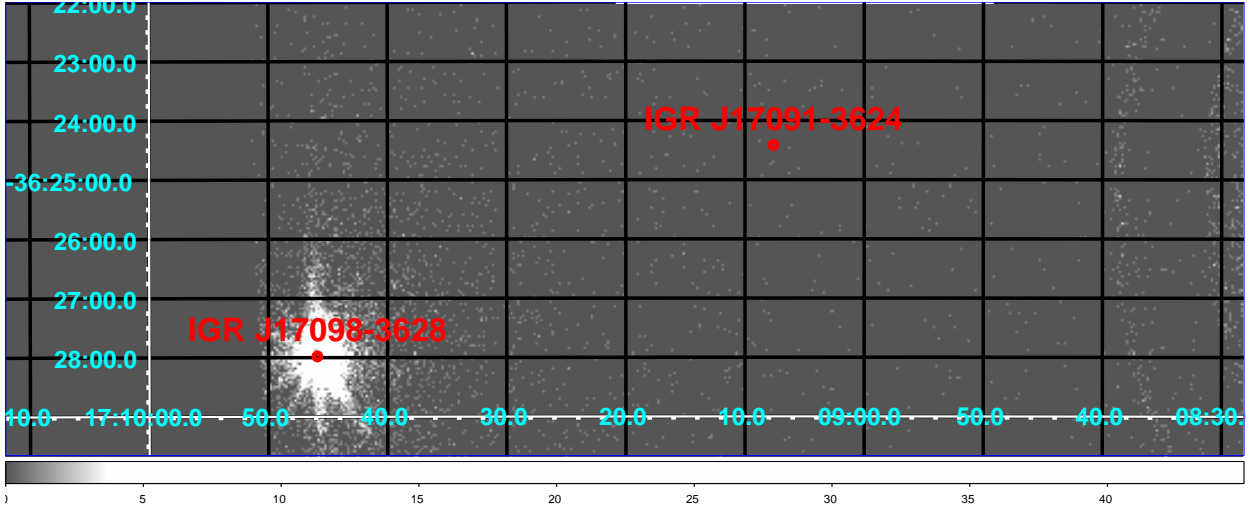}
      \caption{350 s {\it Swift}/XRT row image during 2005. The observation was pointed on IGR J17091--3624, which is not detected. IGR J17098--3628 is quite bright and affected by pileup.}
         \label{2005Swift}
 \end{figure}

The 2005 {\it Swift} observations were performed from May 1 until September 24,
after the peak of the IGR J17098--3628 outburst. A detailed analysis of this
outburst peak has been reported by \citet{Greb2007}. We report here the
spectral analysis of 6 pointings (about 2 ks each) that were performed in
Window Timing (WT).  Figure~\ref{2005Swift} shows the image of one XRT pointing
observation performed in photon counting mode in order to detect the refined
position of the source; as the Figure shows, IGR J17098--3628 is quite bright
while IGR J17091--3624 is not visible.

During all the {\it Swift} observations, IGR J17098--3628 showed a slightly
variable flux of about 3\% with a constant spectral shape. The best fit for
each individual observation is a multicolor disc black body component [{\it
diskbb} in XSPEC ~\cite{diskbb}] with an internal temperature of about 0.9 keV;
the equivalent hydrogen column [$N_H=(0.88 \pm 0.02) \times 10^{22}$~cm$^{-2}$] is
consistent with the galactic one. Table~\ref{2005_par} shows the spectral
parameters of the fit for each observation.
 
The {\it INTEGRAL}/IBIS analysis of the data collected in the same period of
the {\it Swift} observation campaign did not reveal any source emission above
15 keV up to 3 $\sigma$ level. The ASM monitor detected the source with an average
flux of about 30 mCrab, that is a value slightly above the detection limit of
JEM-X monitor (about 20-25 mCrab). Moreover, the source has been observed only
in the JEM-X partially coded field of view because of dithering strategy
used for {\it INTEGRAL} observations. For these reasons, the X-ray monitor
JEM-X did not detect the source in each single science window at the 3 $\sigma$
level. However, adding the JEM-X science windows together yields a total
exposure time of 450 ks which then provides a 36$\sigma$ level detection of
the source in the mosaic.


\begin{table}
\begin{center}
\caption{XRT fit parameters of the 2005 IGR J17098--3628 campaign (errors are at 90\% confidence level):  $T_{in}$= inner temperature of the accretion disk, $N_{disk}$= normalization constant of the {\it diskbb} model, Flux$_{0.5-10}$= unabsorbed model flux between 0.5-10 keV, ${\chi}^{2}_{red}$= reduced $\chi$ square}
\label{2005_par}
\begin{tabular}{lccccc}
\hline
\hline
pointing & Esposure & $T_{in}$ & $N_{disk}$ & Flux$_{0.5-10}$ & ${\chi}^{2}_{red}$ \\
Date & ks & keV & - &ergs  cm$^{-2}$ s$^{-1}$ & -\\
2005-05-01 &0.4 & 1.09$^{0.01}_{-0.01} $ & 76$^{4}_{-4}$ & 2.4$\times$10$^{-9}$ & 1.08 \\
2005-05-10 &1.0 & 1.06$^{0.02}_{-0.02} $ & 80$^{6}_{-6}$ & 1.2$\times$10$^{-9}$ & 0.98 \\
2005-05-21 &1.0 &0.98$^{0.02}_{-0.02}$ &87$^{8}_{-7}$ & 1.6$\times$10$^{-9}$ & 0.99\\
2005-08-13&1.5 &0.82$^{0.02}_{-0.02}$&95$^{11}_{-10}$ & 9.6$\times$10$^{-10}$ & 1.28 \\
2005-09-05& 2.1&0.93$^{0.01}_{-0.01}$&68$^{5}_{-5}$ &1.1$\times$10$^{-9}$& 1.23 \\
2005-09-14&3.2 &0.93$^{0.01}_{-0.01}$&73$^{4}_{-4}$ &1.5$\times$10$^{-9}$& 1.20 \\
\hline
\end{tabular}
\end{center}

\end{table}

\subsection{XMM 2006-2007 Observations}

{\it XMM} observations of the field containing both sources were performed in
2006 August 29 and in 2007 February 19. These observations revealed that IGR
J17091--3624 still did not show any detectable emission, while IGR J17098--3628
was in a relatively bright state.  {\it INTEGRAL} observed the field containing
the two sources at the same time as {\it XMM}, with no detections from either
sources; the {\it INTEGRAL} upper limit flux for a single pointing (2400 s) is
about 20 mCrab in the range 20--100 keV.

\begin{figure}
\plotone{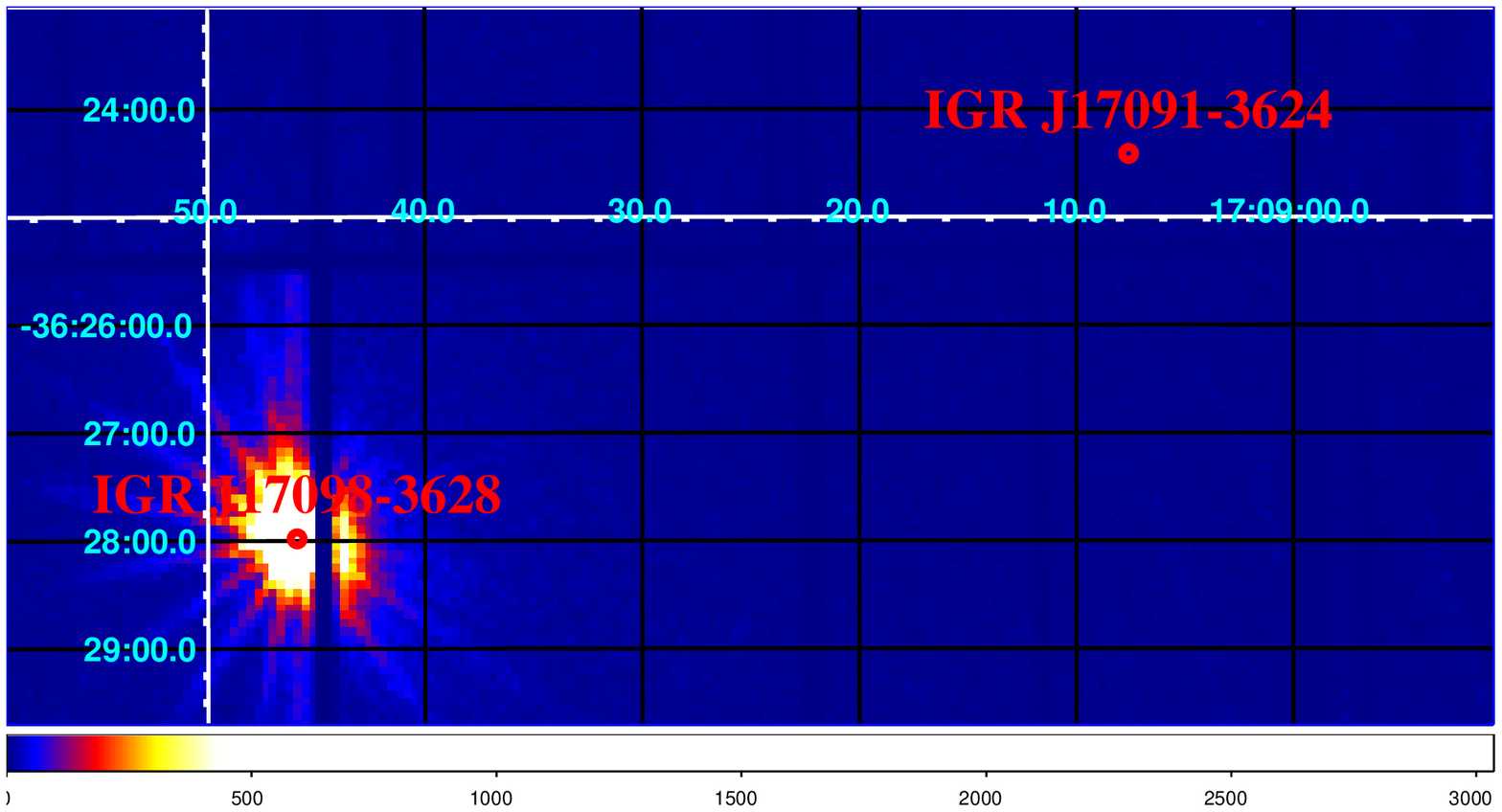}
      \caption{{\it XMM}/EPN image (16 ks) of the 2007-02-19 pointing observation. IGR J17091-3624 is not visible while IGR J17098-3628 is bright and affected by pile up (The dark bar in the image is due to the gap between two adjoining CCDs).}
         \label{2007XMM}
 \end{figure}
 
\subsubsection{IGR J17091--3624}

No detectable emission was revealed from IGR J17091--3624.  Since the distance
of this source is still unknown, we assumed IGR J17091--3624 as located at the
Galactic Center (8 kpc). The corresponding best upper limit luminosity, derived
for the quiescent state of this source and based on the {\it XMM} observations (exposure time 7 ks),
is $L \leq 7 \times 10^{32}$~ergs$^{-1}$ \citep{XMM_s}.

\subsubsection{IGR J17098--3628}

\begin{figure}
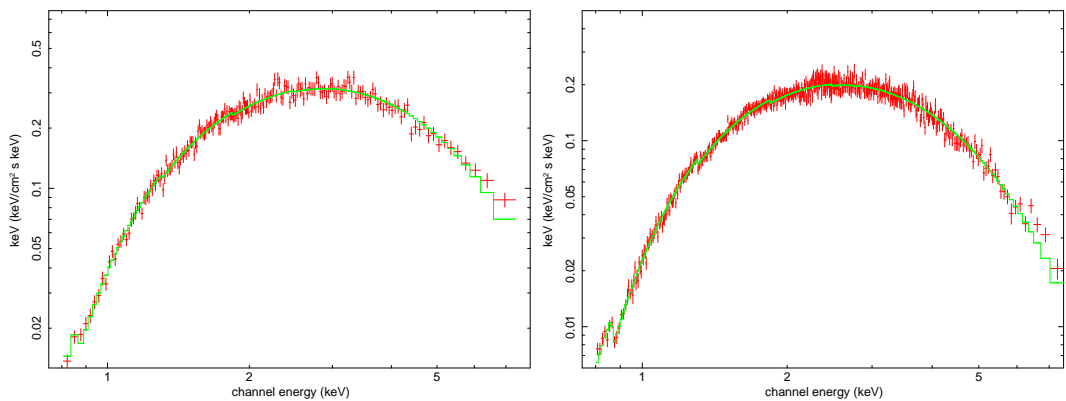

   \centering
 \includegraphics[angle=-90, scale=0.3]{fig5a.eps}
 \includegraphics[angle=-90, scale=0.3]{fig5b.eps}

      \caption{{\it XMM}/PN unfolded spectra of IGR J17098--3628. left panel: 2006-08-25 spectrum. Right panel: 2007-02-19 spectrum. The data are fitted with {\it diskbb} model. Fit parameter values are reported in  Table~\ref{XMMtable}}
         \label{XMMspec}
 \end{figure}

During August 2006 and February 2007, the {\it XMM} spectral shape of the
source did not show, once more, any important variation. The only noticeable
difference was the flux decrease of about 20\% between the two pointings.  Both
spectra are characterized by an absorbed multicolor disc black body emission
with an inner temperature that varied from about 1 keV in the first pointing to  
about 0.9 keV in the second pointing (see Table~\ref{XMMtable}). 
The equivalent hydrogen column, $N_{H}=0.8 \times 10^{22}$~cm$^{-2}$ is in good agreement with
that measured by XRT in 2005 and with the galactic value. Table~\ref{XMMtable}
shows the fit parameters of both pointings. There is no evidence in the
spectral shape of a reflection component or Fe emission line according with the
2005 XRT data set analysis.  Figure~\ref{XMMspec} shows the unfolded spectrum
of the two {\it XMM}/PN pointings.

 \begin{table}
\begin{center}
\caption{PN fit parameters of the two {\it XMM} pointings (errors are at 90\% confidence level): $T_{in}$= inner temperature of the accretion disk, $N_{disk}$= normalization constant of the {\it diskbb} model, Flux$_{0.5-10}$= unabsorbed model flux
between 0.5-10 keV, ${\chi}^{2}_{red}$= reduced $\chi$ square
}
\label{XMMtable}
\begin{tabular}{lccccc}
\hline
\hline
pointing & Exposure &$T_{in}$ & $N_{disk}$ & $F_{0.5-10}$ & ${\chi}^{2}_{red}$ \\
Date & ks & keV & \nodata &ergs~cm$^{-2}$ s$^{-1}$ & \nodata \\
2006-08-25&8 & 1.04$^{+0.01}_{-0.01}$ & 47$^{+3}_{-3}$& 1.2$\times$10$^{-9}$ & 0.85 \\
2007-02-19& 16 &0.87$^{+0.01}_{-0.01}$ &65$^{+2}_{-2}$ & 8$\times$10$^{-10}$ & 0.97 \\
\hline
\end{tabular}
\end{center}
\end{table}

\begin{figure}
\plotone{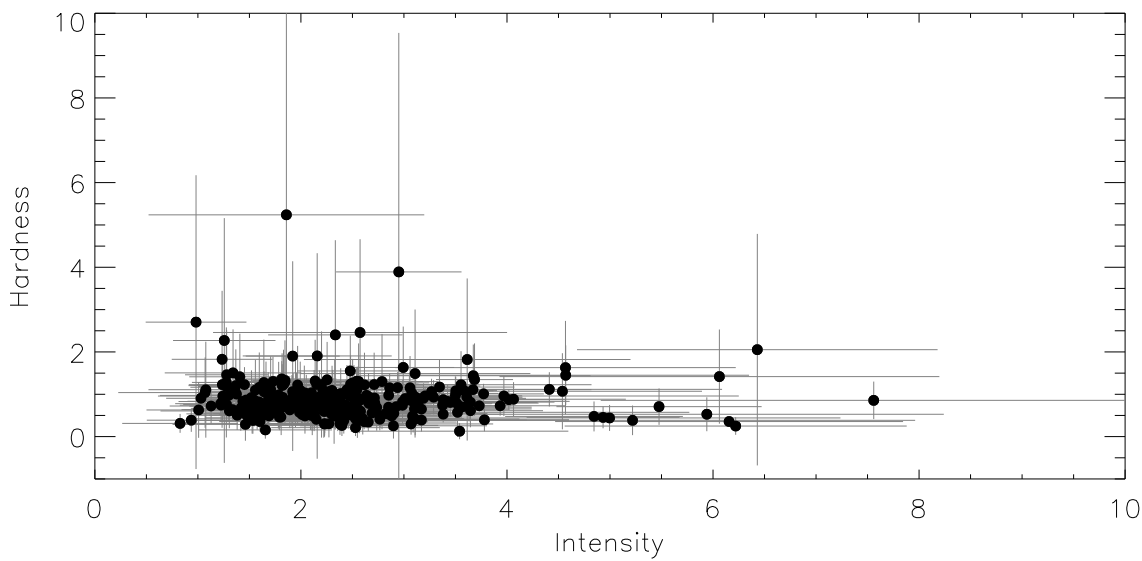}
     \caption{{\it RXTE}/ASM hardness-intensity diagram of the period from 2005 May 1 until 2007 April 10. In this period IGR J17091--3624 was in quiescence and the ASM flux emission is due only to IGR J17098--3628. The hardness ratio is defined as: $HR=({flux_{B}/flux_{A}})$ where A and B are respectively the 1.5--3 keV and 3--5 keV energy bands. The plot clearly shows that the hardness value have not changed all over the observation period.}
          \label{2005Camp}
  \end{figure}

This behavior is confirmed by the {\it RXTE}/ASM light curve from 2005 May 1 to
2007 April 10. In this period, IGR J17091--3624 was in quiescence and the ASM
light curve is only due to IGR J17098--3628 flux emission. As
Figure~\ref{2005Camp} shows, the {\it RXTE}/ASM hardness-intensity diagram
confirms that the source varied only the flux with a constant hardness ratio.

 \subsection{The July 2007 {\it Swift}/XRT monitoring: the new IGR J17091--3624 outburst}\label{2007sec}

\begin{figure}
\plotone{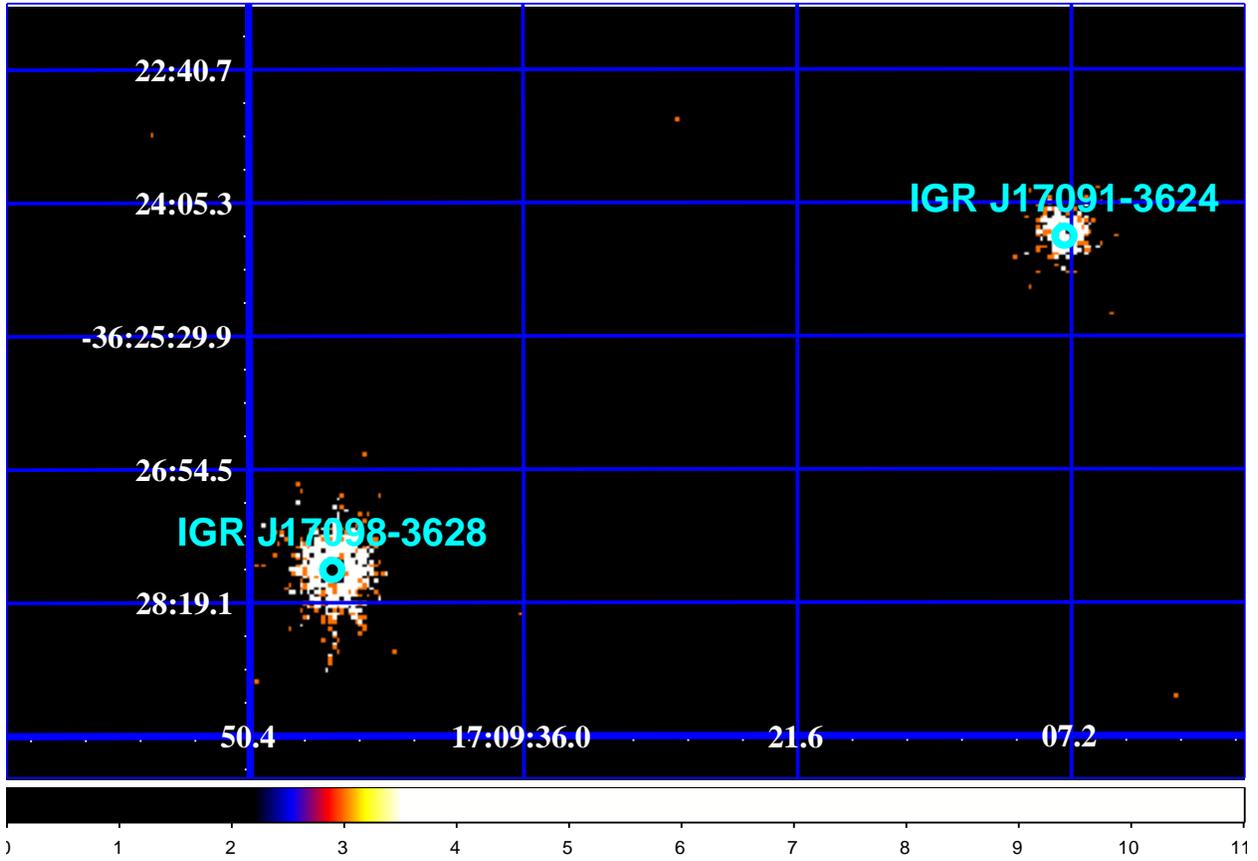}
     \caption{{\it Swift}/XRT 0.1-10 keV image of the July 2007 ToO {\it Swift}  500s observation. The observation has been performed in photon counting mode in order to locate the position of the sources. IGR J17098--3628 is affected by pile up.}
          \label{2007XRT}
  \end{figure}

On 2007 July 9, as a consequence of an increase in the {\it Swift}/BAT and {\it
RXTE}/ASM light curves, a {\it Swift}/XRT 2.5 ks ToO was granted. The
observation was performed in window timing mode in order to avoid the pileup
and revealed the presence of two sources in the XRT field of view, both of
which were causing the flux increase in the RXTE/PCA light curve. Another 500 s
pointing observation, performed in photon counting mode in order to locate the
position of the two sources, revealed the presence of both IGR J17098--3628
and IGR J17091--3624, which  was newly in outburst. Both sources were in
relatively bright state as Figure~\ref{2007XRT} shows.

A {\it Swift} monitoring campaign was then requested for a total of 3 more
observations, 2 ks each (one per week between 2007 July 24 and 2007 August 7)
pointed on IGR J17091--3624. The key outcomes of this observation campaign for
each source are summarized in the following sections.


\subsubsection{IGR J17098--3628}

\begin{figure}
  \centering
     \includegraphics[angle=-90, scale=0.4]{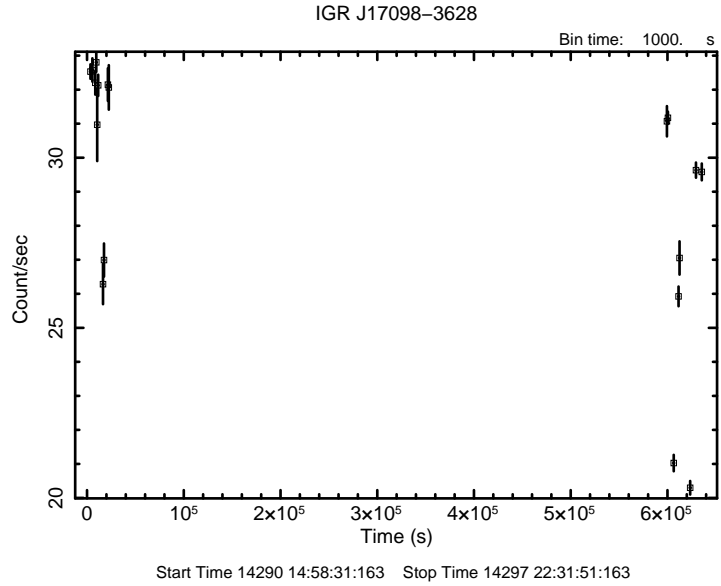}
     \caption{{\it Swift}/XRT 0.1-10 keV light curve of IGR J17098--3628 from 2007-07-09 to 2007-07-17 (bin time: 1000 s). The gap in the light curve is due to the time interval between the second and the third XRT pointing}   
          \label{lc0982007}
  \end{figure}

IGR J17098--3628 has been observed during only three of the pointings associated with the observational campaign, for a total exposure of $\sim$ 8 ks. The source flux only showed flux variations of about 30\% in the range 0.5-10 keV as was inferred from the light curve shown in Figure~\ref{lc0982007}.
The source spectrum revealed only minor changes and the best overall fit is represented by a {\it diskbb} component with an inner temperature of about 0.9-1 keV.

\subsubsection{IGR J17091--3624} 

\begin{figure}
   \centering
 \includegraphics[angle=-90, scale=0.4]{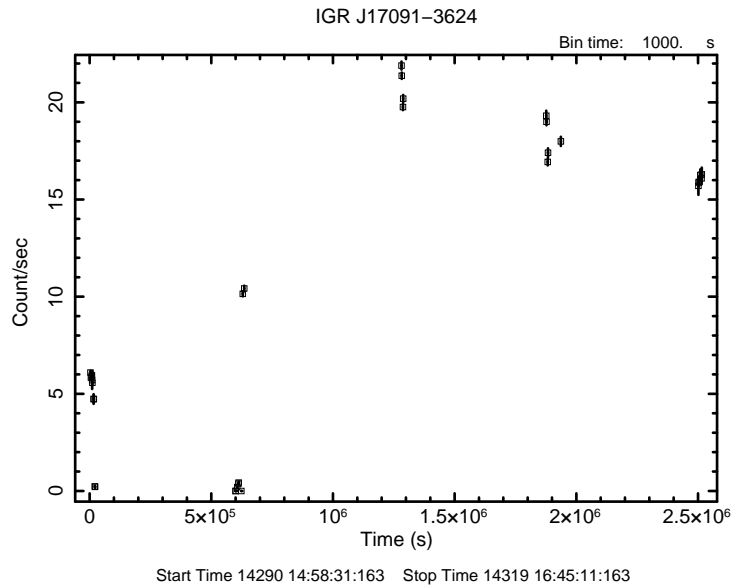}
      \caption{{\it Swift}/XRT (0.1-10 keV) light curve of IGR J17091--3624 from 2007-07-09 to 2007-08-07 (bin time: 1000 s). }
         \label{lcurve_091}
 \end{figure}

This XRT ToO was a unique opportunity to catch the new outburst of IGR
J17091--3624 from the very beginning and also provided the refined position of
the source \citep{KenCap}: RA 17$^h$ 09$^m$ 07.6$^s$, Dec --36$^\circ$ 24$\arcmin$ 24.9$\arcsec$
(J2000), with an estimated uncertainty radius of 3.6$\arcsec$ (90\%
confidence), consistent with the {\it INTEGRAL} position.  However, this X-ray
position rules out the tentative radio counterpart previously proposed for IGR
J17091--3624 \citep{Atel490,Pand}, which lies outside the XRT error circle,
approximately 86$\arcsec$ away.  This location also rules out the optical
counterpart suggested by Negueruela \& Schurch (2006) which was also based on
the radio counterpart.  During the {\it Swift} observations the 0.5-10 keV flux
of IGR J17091--3624 increased from $5 \times 10^{-10}$ (ergs~cm$^{-2}$
s$^{-1}$) to $2 \times 10^{-9}$ (ergs~cm$^{-2}$
s$^{-1}$). Figure~\ref{lcurve_091} shows the 0.1--10 keV light curve of the
2007 XRT monitoring campaign.
 
\begin{figure}
\plotone{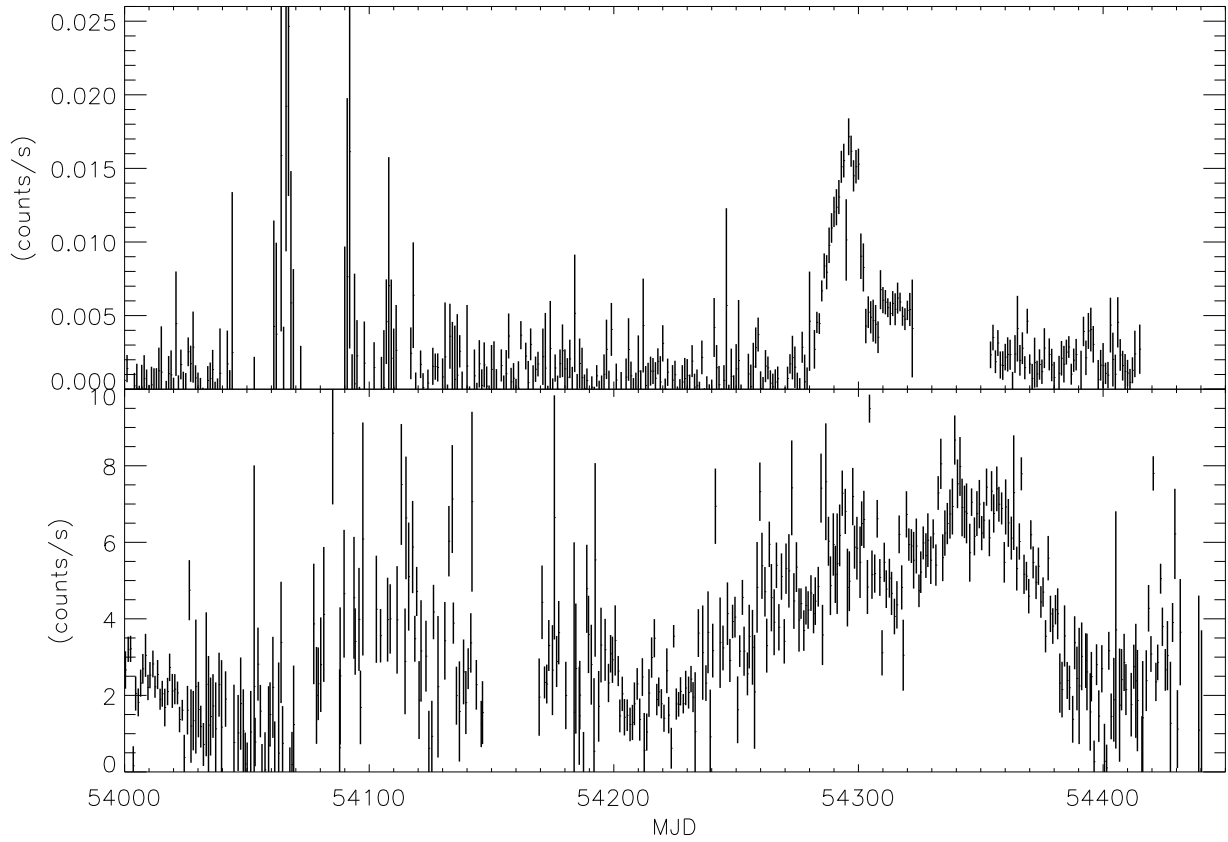}
      \caption{Top panel: IGR J17091--3624 {\it Swift}/BAT (15--50 keV) light curve. Bottom panel: IGR J17091--3624 {\it RXTE}/ASM (2--10 keV). Both curves are contaminated by the IGR J17098--3628 flux emission. The time is expressed in Modified Julian Date (54000 (MJD)= 2006-09-22;54200 (MJD)= 2007-04-10; 54400 (MJD)= 2007-10-27)}
         \label{2007ASM_BAT}
 \end{figure}

In Figure ~\ref{2007ASM_BAT}, we show respectively the {\it Swift}/BAT 15--50
keV (top panel) and {\it RXTE}/ASM 2--10 keV (bottom panel) light curves at the
time of the 2007 outburst.  The luminosity peak present in both light curves is
considered to be mainly due to the IGR J17091--3634 outburst, since the
contamination from the IGR J17098--3628 flux should not imply such a strong
variation of the flux.  In fact, as discussed before, IGR J17098--3628 showed a
soft spectrum and a flux variation clearly lower than the one reported by the
two light curves of Figure~\ref{2007ASM_BAT}.

However, the BAT and ASM light curve show a behavior typical of that expected
at the beginning of a black hole binary outburst, with the hard luminosity peak
preceding the soft one. This behavior is also confirmed by the spectral
analysis: a power law component with a photon index of $1.4 \pm 0.1$ represents
the best fit of the first two observations (9 July and 16 July 2007). 
For the spectra of the subsequent observations, it is necessary to add a
multicolor disc black body component to the power law. Table~\ref{2007_results}
summarizes the spectral parameter values of the XRT 2007 observation campaign.

\begin{table}
\begin{center}
\caption{The fit parameters of the observed spectra of IGR J17091--3624 during 2007 {\it Swift}/XRT monitor campaign (errors are at 90\% confidence level): $T_{in}$= disk inner temperature, $N_{disk}$= {\it diskbb} model normalization constant, $\Gamma$= power law photon index, $N_{pow}$= power law model normalization costant, $Flux_{0.5-10}$= unabsorbed model flux between 0.5-10 keV, ${\chi}^{2}_{red}$= reduced $\chi$ square.}
\label{2007_results}
\begin{tabular}{lcccccccc}
\hline
\hline
pointing & Exposure & $T_{in}$ & $N_{disk}$ & $\Gamma$ & $N_{pow}$ & $Flux_{0.5-10}$ & ${\chi}^{2}_{red}$. \\
Date & ks & keV & \nodata & \nodata &pho~keV$^{-1}$~cm$^{-2}$~s$^{-1}$ &ergs~cm$^{-2}$~s$^{-1}$ & \nodata \\
2007-07-09& 2.4 & \nodata  & \nodata & 1.4 $^{0.1}_{-0.1}$& 0.05$^{0.01}_{-0.01}$ & 4.7$\times$10$^{-10}$ & 0.96 \\
2007-07-16& 3.4 & \nodata & \nodata &1.3$^{0.1}_{-0.1}$&0.028$^{0.003}_{-0.003}$ &  3.0$\times$10$^{-10}$ & 1.10 \\
2007-07-24& 2.6 &0.86$^{0.03}_{-0.03}$ &69$^{13}_{-13}$ &1.6$^{0.1}_{-0.1}$&0.12$^{0.02}_{-0.02}$ & 1.8$\times$10$^{-9}$ & 1.20 \\
2007-08-07& 1.9 & 0.75$^{0.05}_{-0.07}$ &57$^{30}_{-23}$ &2.0$^{0.1}_{-0.1}$&0.23$^{0.04}_{-0.03}$ & 2.1$\times$10$^{-9}$ & 1.08\\
\hline
\end{tabular}
\end{center}

\end{table} 

\begin{figure}
\plotone{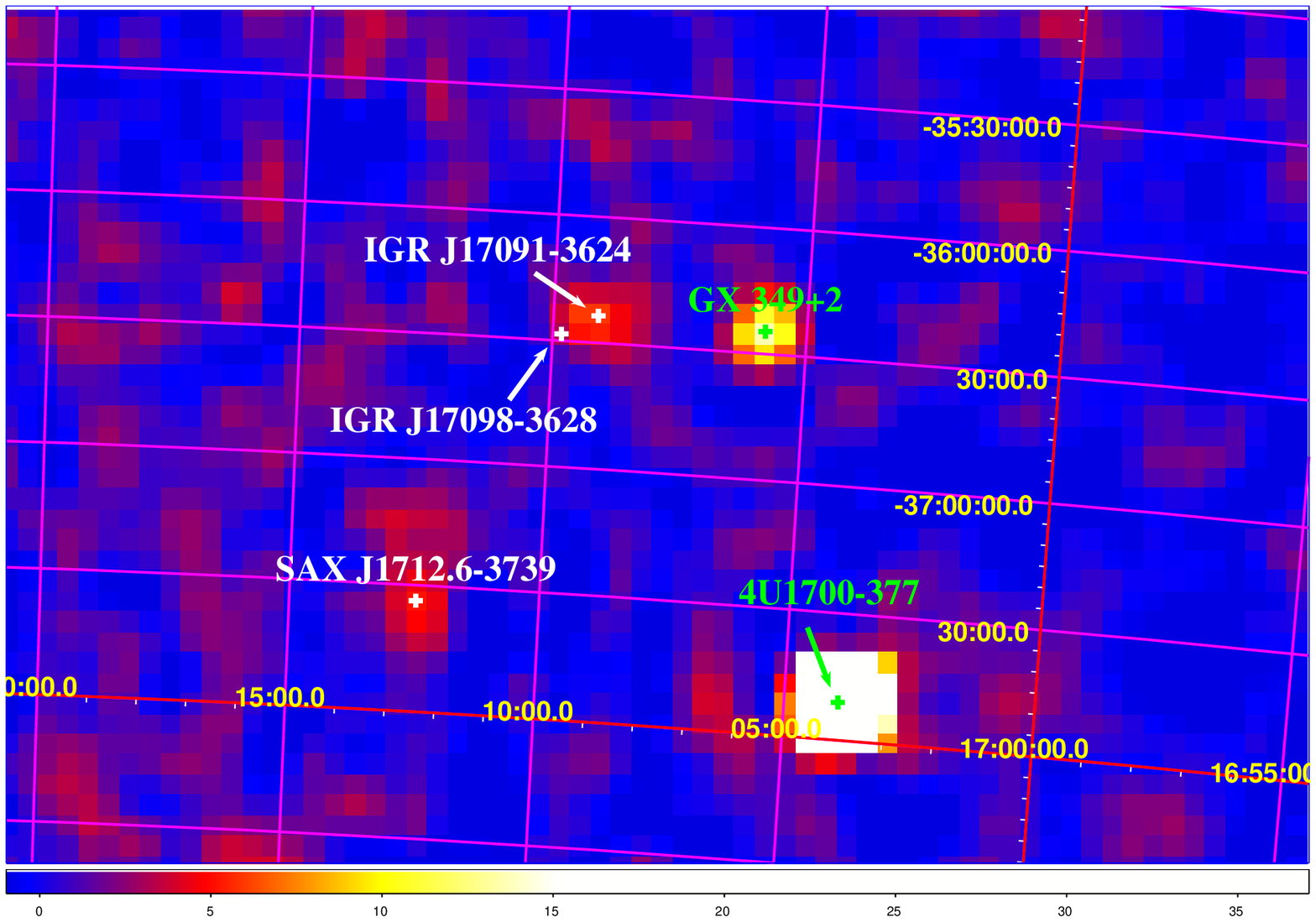}
      \caption{IGR J17091--3624 {\it INTEGRAL}/IBIS 20-100 keV image between August and September 2007 (total exposure time of $\sim$ 50 ks).}
         \label{091_2007}
 \end{figure}

{\it INTEGRAL} detected this source only at the end of the outburst because of visibility constrains. Unfortunately the source was outside the field of view of the {\it INTEGRAL} X-ray monitor JEM-X, thus only the IBIS high energy data were available. 
Figure~\ref{091_2007} shows the 20-100 keV IBIS image of the field of view which includes the  two sources between August and September 2007 (for a total exposure time of $\sim$50 ks). As Figure~\ref{091_2007} shows IBIS detected above 20 keV only IGR J17091--3624. 

\begin{figure}
   \centering
      \includegraphics[angle=-90, scale=0.4]{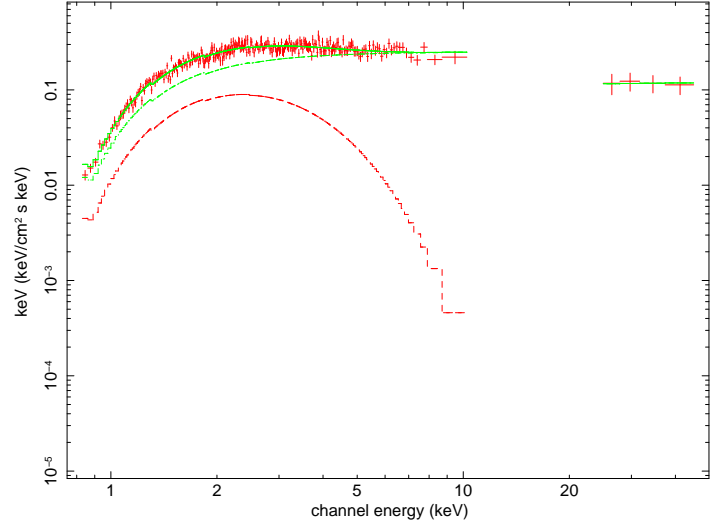}
      \caption{IGR J17091--3624 2007 July outburst: XRT-IBIS combined spectrum: the XRT observation (2 ks) had been performed 18 days before that of IBIS (14 ks). The best fit model is represented by a multicolor disc black body (red dashed line) plus a power law component (green dashed line). The total model is represented in the figure by the solid green line. The spectral parameters are: $T_{in}=0.75 \pm 0.1 $keV, $N_{disk}=57 \pm 25$, $\Gamma =2 \pm 0.1$. The flux between 0.5-100 keV is :  9$\times$10$^{-9}$ ergs  cm$^{-2}$ s$^{-1}$}
         \label{2007ibis_xrt}
 \end{figure}

The first {\it INTEGRAL} IGR J17091--3624 observation of the two sources field (2007-08-25) was performed 18 days after the last XRT observation (2007-08-07). However, as Figure~\ref{2007ibis_xrt} shows, XRT  and IBIS spectra are in good agreement. The {\it INTEGRAL} observations, after the end of the XRT campaign, revealed that the spectrum of the source continued its softening with a photon index that varied from 2 to $\simeq$ 3 at the end of September 2007.  

\subsection{Radio Counterparts} 

\begin{figure}
\plotone{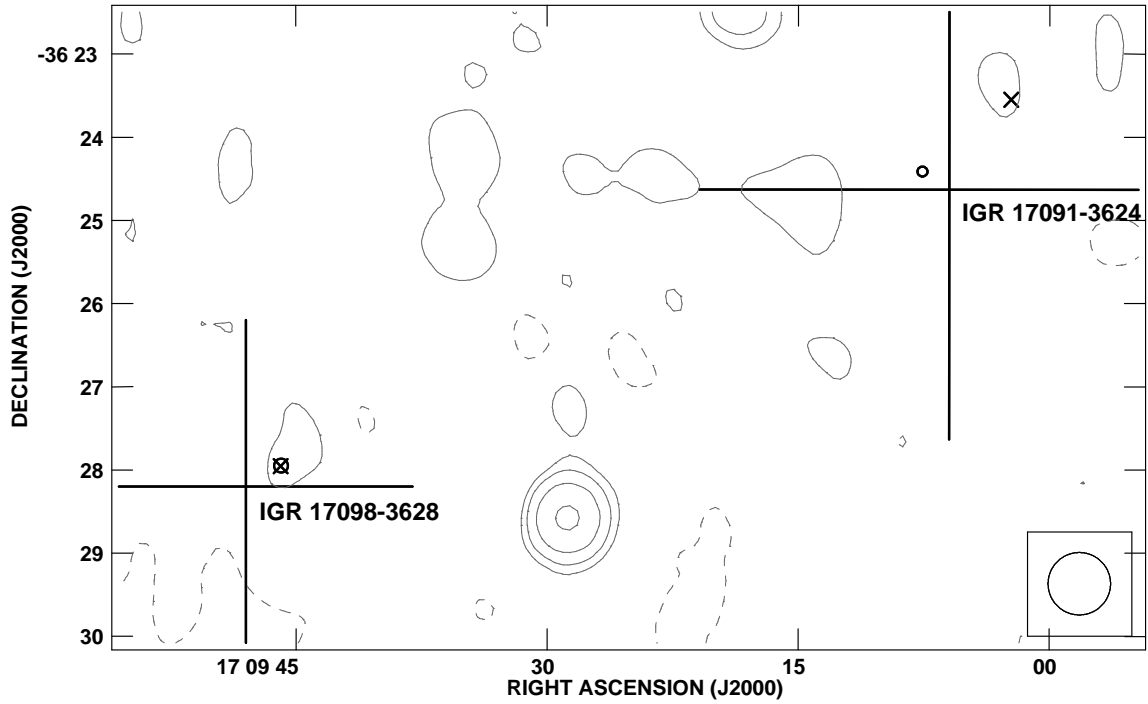}
\caption{NVSS image of the region of IGR J17091--3624 and IGR 
17098--3628, contours traced at ($-1, 1, 2, 4, ...) \times 1$ mJy 
beam$^{-1}$ (thin grey lines). Darker bold symbols represent IBIS 
positions (big plus signs), {\it Swift}/XRT positions (circles, 90\% 
confidence), and previously proposed radio counterparts (crosses). The 
key in the bottom right corner shows the radio beam size ($45\arcsec 
\times 45\arcsec$ HPBM).}
\label{f.nvss}
\end{figure}

A plot of the NVSS data in the region of IGR J17091--3624 and IGR J17098--3628 is
shown in Figure~\ref{f.nvss} together with the position and associated
uncertainties of the {\it INTEGRAL} and {\it Swift}/XRT measurements (marked by
vertical and diagonal crosses, respectively) and the tentative radio
counterparts from VLA literature data \citep{Atel152,Atel490}.  In the region
of IGR J17098--3628, there is clearly an excellent agreement between the radio
and X-ray positions, which gives a high confidence identification. The NVSS
contours themselves show some emission in this region, although it is quite
extended and only significant at the 2.5 $\sigma$ level. In the region of IGR
17091--3624, the refined {\it Swift}/XRT position is clearly not consistent
with the proposed radio counterpart. No signal is detected in the NVSS image in
that region.

\begin{figure}
\plotone{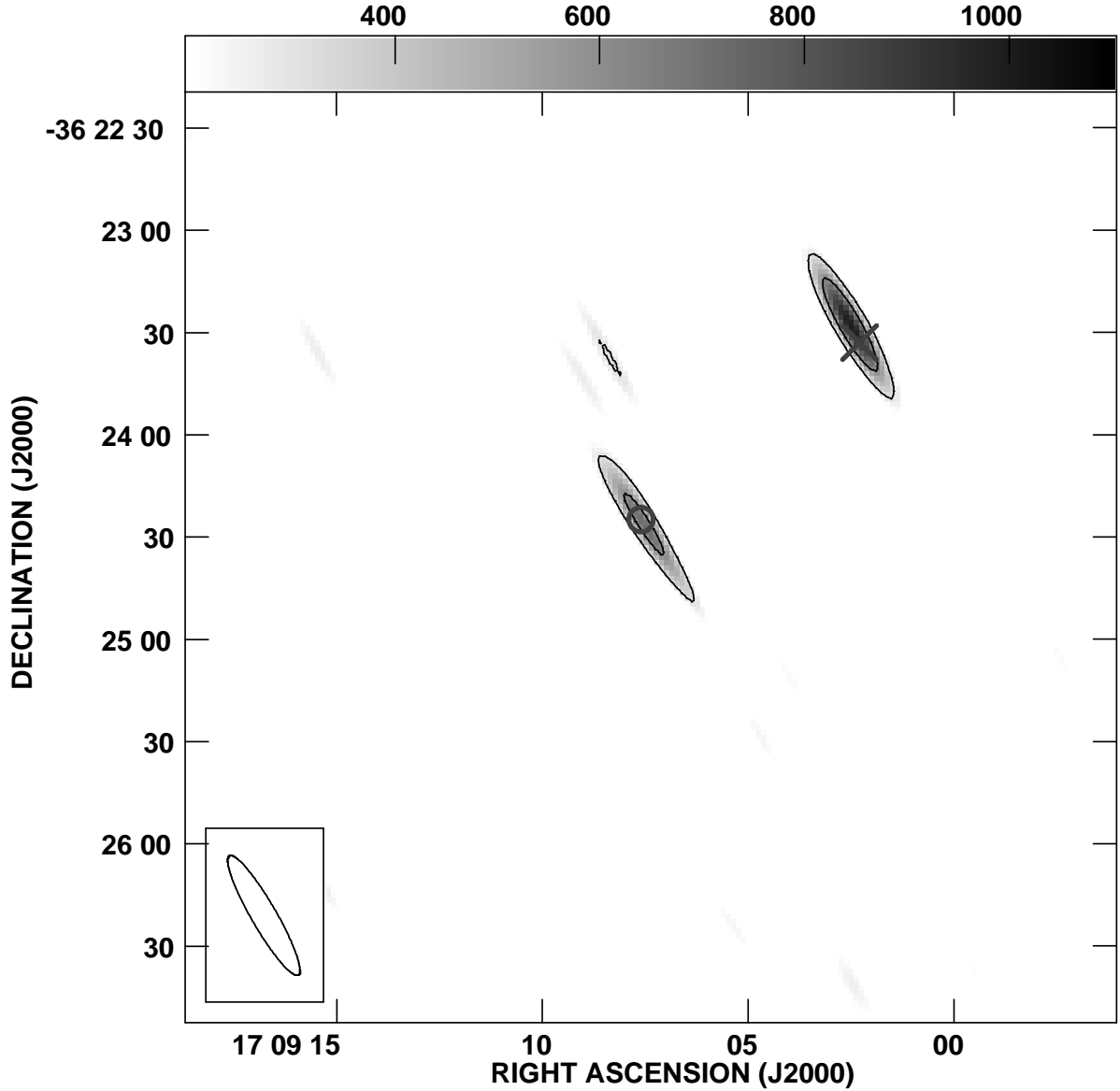}
\caption{Very Large Array 8.4 GHz image of the region of IGR 
17091--3624, taken on 2003 May 09. Contours are traced at $(-0.3, 0.3, 
0.6, 1.2)$ mJy beam$^{-1}$; the gray scale range is 0.2 to 1.1 mJy 
beam$^{-1}$. The circle shows the refined {\it Swift}  position (90\% 
confidence), and the cross shows the previously proposed radio 
counterpart. The key in the bottom left corner shows the radio beam size 
($40.4\arcsec \times 7.3\arcsec$ HPBM in PA $30^\circ$, measured north 
to east).}
\label{f.091}
\end{figure}

We have therefore searched the VLA archive for higher resolution images of
these regions, primarily to find a radio counterpart to IGR J17091--3624 but
also in order to study the time and the spectral evolution of the sources. We
have found archival data from 2003 in the region of IGR J17091--3624 and from
2005 in the region of IGR J17098--3628. The data from 2003 were taken in the
compact D configuration on Apr 23, Apr 26, May 06, and May 09, with exposures
of a few minutes. We show an 8.4 GHz image of the IGR J17091--3624 region in
Figure~\ref{f.091}, superimposed to the {\it Swift}/XRT position. A radio source
is clearly revealed in a position consistent with the high energy error
circle. This component is present in the three epochs at 8.4 GHz, and shows an
increase in total flux. Data at 5 GHz are also available at three epochs; the
source is marginally detected in the first one (3$\sigma$, 5 GHz data only),
and clearly revealed in the next two. For the epochs where simultaneous data
are available, a comparison to the 8.4 GHz flux reveals an inverted
spectrum. Table~\ref{t.091} reports on the measured flux densities; the
position of the radio source is RA 17$^h$ 09$^m$ 07.5$^s$ $\pm$ 0.5$^s$, Dec
--36$^\circ$ 24$\arcmin$ 24$\arcsec$ $\pm$ 3$\arcsec$.

\begin{table}
\begin{center}
\caption{Flux density of the radio counterpart of IGR J17091--3624. The observation of 2003 Apr 23 was at 5 GHz only and is a 3 $\sigma$ detection; 2003 May 09 is only at 8.4 GHz. Spectral index is defined according to $S(\nu) \propto \nu^{-\alpha}$.}
\label{t.091}
\begin{tabular}{lcccc}
\hline
\hline
Date & $S_{\mathrm{5\, GHz}}$ & $S_{\mathrm{8.4\, GHz}}$ & Spectral index \\
 - & (mJy) & (mJy) & -\\
2003 Apr 23 & 0.32 & \nodata & \nodata \\
2003 Apr 26 & 0.48 & 0.56 & $-0.3$ \\
2003 May 06 & 0.42 & 0.67 & $-0.8$ \\
2003 May 09 & \nodata & 0.73 & \nodata \\

\hline
\end{tabular}
\end{center}
 
\end{table} 

\section{Discussion and Conclusions}  \label{disc}
The huge amount of data collected and the broad energy coverage allowed us to perform a detailed spectral analysis of these two sources, following their evolution during more than two years and fixing upper limits when one of the sources was not visible. In the following a detailed summary of the results for both sources can be found.
 
\subsection{IGR J17098--3628} 
Our analysis, spanning from May 2005 to September 2007, proves that the source
spectrum shows a soft black body component with an internal temperature of
about $\sim$ 1 keV and an internal radius comparable to the last stable orbit
of the accretion disc. The luminosity due to soft component \citep[assuming
$d=10.5$ kpc,][]{Greb2007} varied during the two years of our observational
campaign from $\sim 2 \times 10^{37}$ to $\sim 5 \times 10^{36}$
ergs~s$^{-1}$. The flux variation range and the spectral parameters are
comparable with the ones reported by \citet{Greb2007} during the first phases
of the outburst on April 2005.  In our data set each flux variation seems not
to be correlated with any relevant spectral features, as confirmed by the lack
of evident variation of the {\it RXTE}/ASM hardness-intensity diagram
(Figure~\ref{2005Camp}).  However respect to the previous April 2005
observations, we did not detect any power law emission at higher {\it
INTEGRAL}/IBIS energy band. This indicates that the spectral shape was evolved
from the first phase of the outburst being dominated only by a thermal disc
component coming from the accretion disc around the compact object.

We can conclude that this source  seems  to have spent 2.5 years in the high soft state with a disc black body component substantially identical to the one previously observed at the beginning of the outburst~\citep{Greb2007}. Curiously the only difference is the lack of any high energy emission. In fact the hard component fell below the detection limit of IBIS after about three months from the beginning of the outburst. Hence the geometry and the temperature of the accretion disc have not shown any significant variation up to now, on the other end the power law emission quenched probably because of the electron temperature of the corona fallen below the disc seed photons temperature making the inverse Compton scattering processes inefficient.

\subsection{IGR J17091--3624} 

Previous studies have demonstrated that the X-ray luminosity of the BHCs in quiescence is lower than that of neutron star X-ray binaries and falls below $\sim$ 10$^{32}$-10$^{33}$ ergs~s$^{-1}$ \citep{Campana}. In the quiescent state, neutron star X-ray binaries are normally detected and hence the {\it XMM} upper limit of IGR J17091--3624 is a good indication that the source is a BHC. 

 The {\it Swift}  ToO caught the source at the beginning of its outburst when it was still in a low/hard state. The best fit of the first observation  (2007-07-09/16) is a power law spectrum  with $\Gamma$=1.4$\pm$ 0.1. The source spectrum substantially softened and the black body component became dominant with an increasing temperature and a steeper power law.
{\it INTEGRAL} continued to monitor the source with IBIS  after the end of the XRT observational campaign, up to the end of September 2007. These observations confirm the softening of the source. A power law component without an energy cutoff provided the best fit of the IBIS spectra between 20 and 80 keV, probably due to jets or corona reprocessing of disc seed photons. The power law photon index varied during two month period from about 1.4 (beginning of July) to about 3 (end of September). Unfortunately the relatively low source flux at high energies (7 $\times$ 10$^{-10}$ ergs~cm$^{-2}$ s$^{-1}$ between 20-100 keV) and the variation of the spectral shape did not provide sufficient statistics to extend the spectra up to 80 keV and to verify the presence or not of an high energy cutoff in the spectrum. 
 
 Thanks to the XRT refined position, we found the radio counterpart
of the source. As soon as 9 days after the detection by IBIS in 2003, a radio
source at the sub-mJy level was detected at 5 GHz. The source has been detected
also at 8.4 GHz, and it showed an increase in flux over the subsequent two
weeks. The spectrum is inverted, characteristic of self-absorbed synchrotron
radiation from a compact jet. This behavior is typical of BHC in low hard
states. Although it is difficult to guess the spectral shape at higher
frequencies, it is reasonable to estimate that the total radiative luminosity of
the compact jet is of the order of 10$^{31}$ erg sec$^{-1}$, assuming the distance to
the Galactic Center. No information is available after the 2007 outburst, and it
will be clearly valuable to obtain new radio observations after future episodes
of activity.

Chaty et al. 2008, on the basis of ESO NTT observations, report on two possible infrared counterparts of IGR J17091--3624 consistent with the {\it Swift}/XRT error box. Our refined position, based on the radio observations, could support the identification of the real infrared counterpart of this source.

Little is known about nature, duration and recurrence of outbursts in transient X-ray binaries. It is probable that the outbursts are due to randomly acting factors such as the mass transfer variations or truncation of the inner disc radius. However for IGR J17091--3624 five outbursts are known from 1994 to 2007. Thus, it appears that this source goes in outburst every three or five years. Generally the flux varies from 5 mCrab to about 20 mCrab in the range 2-10 keV over a period of few months.  The 2003 outburst was the first detected  at higher energy range (20-150 keV) for one year period. Comparing the two hard states of respectively the 2003 and 2007 outbursts, the last one seems to be the hardest. In fact it was possible to determine the high energy power law cutoff of the 2003 hard state \citep[$\sim$ 49 keV,][]{Cap2006}  while in the 2007 hard state the cutoff is not detectable up to 80 keV. Concerning the soft state the 2003 outburst has an higher temperature black body emission~\citep{Cap2006} that does not appear to be present in the 2007 outburst. However the XRT monitoring campaign observed the source only at the beginning of its transition to the soft state and unfortunately in the subsequent {\it INTEGRAL} monitoring the source was outside the JEM-X field of view and so it was not possible to follow the entire evolution of the black body temperature. 

\begin{acknowledgements}
This work has been supported by the Italian Space Agency through grants  I/008/07/0 and I/088/06/0. 
We acknowledge the use of public data from the {\it Swift}  data archive and all the {\it Swift} team for its support and Memmo Federici for supervising the {\it {\it INTEGRAL}\/} data analysis. 
Particular thanks are due to Erick Kuulkeers for making his private data available for analysis.
\end{acknowledgements}

\end{document}